\documentstyle[12pt]{article}

\begin{document}

\vspace{1cm}
{\large
Non-rigid Shell Model and Correlational Mechanism of the Local 
Pairing.
}

\vspace{0.5cm}
{\it A.S. Moskvin, V.A. Korotaev, Yu.D. Panov, M.A. Sidorov.}

\vspace{0.3cm}
{\small
Department of Theoretical Physics, Ural State University,
620083, Lenin ave. 51, Ekaterinburg , Russia.
}
\vspace{0.5cm}

{\small
The Hartree-Fock states of the many-electron atomic system can be 
unstable with respect to a static or dynamic shift of the electron 
shells. An appropriate non-rigid shell model for atomic clusters is 
developed. It permits to formulate a convenient approach to the 
semiempirical description of the different correlation effects and to 
reveal some new effects.
}
\vspace{1cm}

In the framework of conventional Hartree-Fock (HF) scheme the 
description of some specific correlation effects in atomic systems 
requires too large set of HF-configurations. This is connected, in 
particular, with the high density of excited states, with possible 
anomalously strong influence of small perturbations, and the 
instability with regard to a formation of strongly correlated states. In 
such situations, as a rule, special model approaches are used.
One of the most simplest descriptions of the correlation effects is 
connected with the taking into account shifts of atomic orbitals of the 
conventional MO-LCAO-scheme [1]. For doubly occupied orbitals 
(biorbitals) with shifted electron shells
$$
\Psi(\vec{r_1},\vec{r_2};\vec{q_1},\vec{q_2}) = 
[2+2S(\vec{q_1} - \vec{q_2})]^{-\frac{1}{2}} 
[\phi(\vec{r_1} - \vec{q_1}) \phi(\vec{r_2} - \vec{q_2}) +
\phi(\vec{r_1} - \vec{q_2}) \phi(\vec{r_2} - \vec{q_1})]
$$
where overlap integral
$$
S(\vec{q_1} - \vec{q_2}) = 
\langle \phi(\vec{r} - \vec{q_1}) | 
\phi(\vec{r} - \vec{q_2}) \rangle .
$$
These states provide an extremum of the full energy functional 
$E[\Psi] = \langle \Psi | H | \Psi \rangle$, where $H$ is usual 
helium atom Hamiltonian, at $\vec{q_1} + \vec{q_2} = 0$, $| 
\vec{q_1} - \vec{q_2} | = q_0$. The «extremal» orientation of 
$\vec{q} = \vec{q_1} - \vec{q_2}$ depends on type of one-particle 
$\phi$-state. In a case of $ns$-states the extremum of the functional 
$E[\Psi] = E(q)$ is realized at any orientation of $\vec{q}$ with 
fixed length $q_0$. For $np$-states the situation is more 
complicated. At $|\phi \rangle = p_z$ there are possible the 
«easy axis»- and «easy plane»-type shifts relative z-axis 
($np_{\sigma}^2$- and $np_{\pi}^2$-configuration respectively). In 
Fig.1 and 2 $E(q)$ are plotted for simplest electronic configurations 
of the helium atom : $1s^2$ (Fig.1; analytic computation), and 
$2p_z^2$  (Fig.2; numerical computation).

Note that in all considered cases there is a minimum of the functional 
$E(\vec{q})$ at $q \neq 0$, i.e. when atomic orbitals are shifted. 
The minimum is weakly prominent for $1s^2$- and $2p_{\pi}^2$-
configurations. On the contrary, in case of the $2p_{\sigma}^2$-
configuration, the depth of minimum at $q \simeq 0.8$ a.u. reaches value 
of some eV.

These results do not contradict to well known ones for the helium 
atom with taking account of the effective charge $Z^{\ast}$ as a 
supplementary variational parameter. A non-trivial 
$E(\vec{q}, Z^{\ast})$   minimum for the $1s^2$-configuration 
with the non-zero shell shift exists only at $Z^{\ast} > Z-\frac{3}{16}$ 
otherwise there is only a trivial minimum at $q=0$. As well known 
an absolute $E[\Psi]$ minimum corresponds to 
$Z^{\ast} = Z-\frac{5}{16}$ and $q=0$.

The $He$ atom example has rather illustrative character. However it 
shows that in crystals where effective one-particle potential being 
beforehand unknown, the one-particle basis of shifted atomic HF-
orbitals permits to describe some correlation effects with a restricted 
set of initial atomic orbitals.

Besides the static shifts of atomic shells, one can consider also 
dynamic shifts related, e.g., to the rotation of an atomic shell about 
the initial center (spontaneous current). In new states electric (dipole, 
quadrupole) or magnetic polarizability may be anomalously large and 
its magnitude being determined by the interelectronic correlation 
interactions.

Within a non-rigid shell model we have to modify a conventional 
LCAO-approach introducing a set of the molecular orbitals based on 
shifted atomic functions characterized by the symmetrized 
$q^{\gamma}$-shift modes in analogy to the appropriate nuclei 
displacements.

A non-rigid shell model gives a simple and obvious example of a 
local pairing within the two-electron $s^2$-like configurations as a 
result of the correlation effects. The local pairing is promoted by the 
presence of a strongly polarized shell, as well as the orbital 
degeneracy or quasi-degeneracy within valent states (for simplicity, 
$d$-states) through the electric multipole $s-d$-interaction described 
by the effective «vibronic-like» Hamiltonian
$$
V^{sd} = \Sigma B_{\gamma}(\hat{V}^{\gamma} q^{\gamma}),
$$
where the $\hat{V}^{\gamma}$-operator acts within $d$-manifold, 
the $B_{\gamma}$ are «vibronic-like» parameters. This interaction 
can result in a purely electronic Jahn-Teller effect.

In general we have to take account of the atomic displacements 
$Q^{\gamma}$-modes and their interaction with electronic 
$q^{\gamma}$-shifts:
$$
V_{qQ} = \Sigma b_{\gamma}(q^{\gamma} Q^{\gamma}).
$$
This results in a complicated multi-mode Jahn-Teller effect with a 
correlational hybridization at the $s$-, $d$-electron modes and the 
local structural modes.

This system will have all anomalous properties of a Jahn-Teller 
center, in particular, large values of the low-frequency polarizability. 
It appears that within the non-rigid shell model the completely filled 
electron shells do not quenched and can reveal many peculiarities 
similar to the nonfilled shells.

The magnitudes of the shell $q^{\gamma}$-shifts are correlation 
parameters which may be found by minimizing the energy functional 
$E(q)$. The quantity $\Delta = E(0) - E(q_0)$ determines the pairing 
energy, i.e., the local boson binding energy.

A non-rigid shell model can be considered as a generalization of the 
well known shell model of the lattice dynamics and of the non-rigid 
anionic background model by J.E.Hirsch et al [2]. In particular a 
correlational pseudospin formalism can be successfully applied for a 
description of the valent states for the atomic systems with a 
correlational near degeneracy.

In  conclusion we would like to conjecture the posible importance of 
the non-rigid shells correlation effects for a local pairing in copper 
oxides.

\vspace{2cm}
REFERENCES
\vspace{0.5cm}

1. W. Cosman. {\it Introduction to Quantum Chemistry.} Moscow, MIR, 
1960 (in Russian).

2. J.E.Hirsch et al. Phys.Rev.B {\bf 40} (1989) 2179.
\end{document}